\documentclass[twocolumn,twoside,slac_two]{revtex4}
\usepackage{graphicx}
\usepackage{fancyhdr}
\usepackage{graphics}
\usepackage{epstopdf}
\usepackage{textpos}
\begin{document}

%%%%%%%%%%%%%%%%%%%%%% WRITE THE TITLE HERE %%%%%%%%%%%%%%%%%%%
\title{\centering LHC Phenomenology}
%%%%%%%%%%%%%%%%%%%%%% WRITE THE AUTHOR HERE %%%%%%%%%%%%%%%%%
% 40+5 min; 7-8 pages for proceedings
% Session convener: David Morrissey (send .pdf for approval by Thurs November 10)

%%% Please insert your personal picture here!

\author{
\centering
\begin{center}
Heather E.\ Logan
\end{center}}
\affiliation{\centering Ottawa-Carleton Institute for Physics, Carleton University, Ottawa K1S 5B6 Canada}
%%%%%%%%%%%%%%%%%%%%%% WRITE THE ABSTRACT HERE %%%%%%%%%%%%%%%%
\begin{abstract}
The analyses of the first 1--2~fb$^{-1}$ of Large Hadron Collider (LHC) data are already having significant impacts on a wide range of models.  In this talk I give my perspective on why we expect to find new physics at the LHC, and how such a discovery might unfold.
\end{abstract}

\maketitle
\thispagestyle{fancy}


\section{INTRODUCTION: WHY WE EXPECT NEW DISCOVERIES AT THE LHC}

The Standard Model (SM) of particle physics is extremely successful to date.  When searching for new phenomena at the LHC, it makes sense to ask the question, ``can't we get by with just the degrees of freedom that we've already observed?''  The SM already contains:\footnote{Dark matter, the baryonic matter-antimatter asymmetry, and a quantum theory of gravity also need to be incorporated into a complete theory of nature, but their dynamics need not have anything to do with the energy scale being probed by the LHC.}
\begin{itemize}
\item three generations of quarks with their mixing described by the CKM matrix;
\item three generations of charged leptons and three generations of massive neutrinos\footnote{Something new is needed here---the two possibilities are Dirac neutrino masses, which require new unobserved right-handed gauge-singlet neutrino states to be added to the SM, or Majorana neutrino masses, which require new dynamics giving rise to a dimension-five Majorana mass operator---however, these options need not give rise to new physics observable at the LHC.} with their mixing described by the PMNS matrix;
\item gluons from the SU(3) strong interaction;
\item the photon plus the massive $W^{\pm}$ and $Z$ bosons from electroweak SU(2)$\times$U(1).
\end{itemize}
The SM Higgs boson is not part of this list since it has not yet been observed.  We know that electroweak symmetry is broken, but can we get by without the underlying dynamics appearing at the LHC?  The answer to this question is \emph{no}: the SM without a Higgs is intrinsically incomplete.

The incompleteness of the electroweak theory without a Higgs (or some other dynamics to play its role) is exposed by studying $2 \to 2$ scattering of longitudinally-polarized $W$ or $Z$ bosons.  These longitudinal polarization states correspond to the three would-be Goldstone bosons of the SM Higgs doublet; without the physical Higgs particle, the scalar part of the theory is a nonlinear sigma model which becomes strongly coupled at an energy not too far above the weak scale, leading to a violation of perturbative unitarity in the scattering amplitudes.  In terms of the longitudinal $W$ and $Z$ bosons, individual Feynman diagrams contributing to the $2 \to 2$ scattering amplitude grow like $g^2E^4/M_W^4$, where $E$ is the incoming gauge boson energy in the centre-of-mass frame.  Relations among the quartic and triple gauge couplings enforced by SU(2)$\times$U(1) ensure that the bad $E^4$ behaviour cancels among the contributing diagrams, leaving a scattering amplitude proportional to $g^2 E^2/M_W^2$ in the absence of a Higgs.  This amplitude violates unitarity at an energy scale $E = \sqrt{8 \pi} v \simeq 1.2$~TeV.  In the absence of a Higgs, this implies that the physical scattering process must unitarize itself through nonperturbative strong dynamics effects---for example, exchange of heavy composite vector resonances---below this energy scale.

The SM Higgs boson cures this problem by contributing additional scattering diagrams involving Higgs boson exchange that exactly cancel the bad $E^2$ behaviour.  The full SM scattering amplitude retains a term proportional to $M_H^2/M_W^2$ in the high-energy limit; again imposing perturbative unitarity leads to an upper bound on the Higgs mass $M_H \lesssim 1.0$~TeV~\cite{LQT}.

The fact of electroweak symmetry breaking thus provides a bulletproof argument for new phenomena at the energy scale being probed\footnote{Whether these phenomena will be detectable at the LHC is a separate question, mostly related to backgrounds and luminosity.} by the LHC: either the scalar(s) of a weakly-coupled Higgs sector, or the resonances of a strongly-coupled TeV-scale theory.

In the rest of this talk I first describe SM Higgs phenomenology and the existing experimental constraints (Sec.~\ref{sec:Higgs}).  Then I review two frameworks for physics beyond the SM: supersymmetry (Sec.~\ref{sec:SUSY}) and composite models (Sec.~\ref{sec:composite}).  I finish with a brief outlook in Sec.~\ref{sec:outlook}.

%%%%%%%%%%%%%%%%%%%%%%%%%%%%%%%%%%%%%%%%%%%%
\section{HIGGS PHENOMENOLOGY AND CONSTRAINTS}
\label{sec:Higgs}

\subsection{Standard Model Higgs Mechanism}

In the SM, electroweak symmetry is broken by an SU(2)-doublet scalar field,
\begin{equation}
	\Phi = \left( \begin{array}{c} G^+ \\ (H + v)/\sqrt{2} + i G^0/\sqrt{2} \end{array} \right).
\end{equation}
Here $G^+$ and $G^0$ are the Goldstone bosons eaten by the $W^+$ and $Z$, $v = 2 M_W/g \simeq 246$~GeV is the SM Higgs vacuum expectation value (vev), and $H$ is the physical SM Higgs boson.

Electroweak symmetry breaking is caused by the Higgs potential, the most general gauge-invariant renormalizable form of which is,
\begin{equation}
	V = \mu^2 \Phi^{\dagger} \Phi + \lambda \left( \Phi^{\dagger} \Phi \right)^2,
	\label{eq:V}
\end{equation}
where $\lambda \sim \mathcal{O}(1)$ and $\mu^2 \sim - \mathcal{O}(M^2_{\rm EW})$.  The negative value of $\mu^2$ leads to a minimum away from zero field value, causing electroweak symmetry breaking.  Minimizing the potential, the Higgs vev is 
\begin{equation}
	v^2 = -\mu^2/\lambda \simeq (246~{\rm GeV})^2,
\end{equation}
and the physical Higgs mass is 
\begin{equation}
	M_H^2 = 2 \lambda v^2 = -2 \mu^2.
	\label{eq:MH}
\end{equation}
The couplings of the physical Higgs boson to other SM particles are predicted entirely in terms of $v$ and the known particle masses via the usual SM Higgs mass generation mechanism.

\subsection{Existing Constraints and LHC Prospects}

Searches for the SM Higgs boson have been performed at LEP, the Tevatron, and the LHC.  To date these exclude various ranges of possible SM Higgs masses.  They also provide constraints on non-SM Higgs bosons with different production cross sections or branching ratios into the searched-for decay modes.

The LEP experiments searched for SM Higgs production via $e^+e^- \to Z^* \to ZH$ with Higgs decays mainly to $b \bar b$ and $\tau\tau$.  They exclude a SM Higgs with mass below 114.4~GeV~\cite{Barate:2003sz}.  A lighter non-SM Higgs is allowed if the $ZZH$ coupling is suppressed relative to its SM value.

The Tevatron experiments are searching for the SM Higgs in a variety of production and decay channels, the most important of which are $gg \to H \to WW$ and associated $WH$ or $ZH$ production with $H \to b \bar b$, $\tau\tau$, and $WW$.  The most recent combination, using up to 8.6~fb$^{-1}$ per experiment, excludes a SM Higgs with mass in the ranges\footnote{The Tevatron SM Higgs analyses did not consider Higgs masses below 100~GeV.} 100--109~GeV and 156--177~GeV~\cite{Tevatron:2011cb}.  Before shutting down in September 2011, the Tevatron experiments collected a total of about 10~fb$^{-1}$ per experiment.  The final Tevatron combined Higgs analysis, expected in 2012, anticipates at least 2$\sigma$ sensitivity to a SM Higgs boson in the mass range 100--185~GeV, with 3$\sigma$ sensitivity for $M_H \simeq 115$~GeV~\cite{TevRun3Case}.

The multipurpose LHC experiments, ATLAS and CMS, are also searching for the Higgs and have provided the most stringent limits to date.  The most constraining channels involve inclusive Higgs production (mostly from gluon fusion) with decays to $WW$, $ZZ$, and $\gamma\gamma$.  The current analyses consider Higgs masses in the range 110--600~GeV.  The most recent analyses, using up to 2.3~fb$^{-1}$ of data at 7~TeV collision centre-of-mass energy, together exclude all SM Higgs masses at 95\% confidence level except for $M_H$ below 145~GeV, between 288 and 296~GeV, and above 464~GeV~\cite{LP11SMHiggs}.\footnote{A non-SM Higgs with a sufficiently suppressed gluon-fusion production cross section and/or suppressed decay branching ratios into $WW$ or $ZZ$ final states is of course still allowed in the mass ranges excluded for the SM Higgs.} 

The ``light'' Higgs mass region between the LEP lower bound of 114.4~GeV and the LHC exclusion starting at 145~GeV is favoured in the SM by precision electroweak constraints: considering only indirect constraints, the precision electroweak fit puts a 95\% confidence level upper bound on the SM Higgs mass of 161~GeV, with the best-fit Higgs mass below the LEP lower bound~\cite{LEPEWWG:summer2011}.\footnote{Heavier Higgs masses can be consistent with the electroweak fit if new, non-SM particles contribute appropriately to the precision observables.}  The ``channels to watch'' in this favoured mass range are $\gamma\gamma$ (especially below 125~GeV), and $WW \to \ell\nu\ell\nu$ and $ZZ \to 4 \ell$ (especially above 120~GeV).  Two additional channels, $H \to \tau\tau$ and associated $WH$ production with $H \to b \bar b$, also contribute weakly to the LHC sensitivity in this mass range.  Analyses of Higgs production via vector boson fusion and via associated production with $t \bar t$ have not yet been completed.

ATLAS and CMS have each now recorded over 5~fb$^{-1}$ of data at 7~TeV.  This should allow 2$\sigma$ (3$\sigma$, 5$\sigma$) sensitivity to a ``light'' SM Higgs boson for masses above the LEP limit (125~GeV, 140~GeV)~\cite{atlasprojections}.

This ``light'' Higgs mass region happens to be the most interesting for testing the SM Higgs mechanism of mass generation by measuring the couplings of the Higgs to SM particles.  Below the $WW$ threshold, the Higgs branching ratios to $b \bar b$, $\tau\tau$, and $\gamma\gamma$ are large enough to be detected.  Also, vector boson fusion and the associated $WH$, $ZH$, and possibly $t \bar t H$ production channels have large enough cross sections to be experimentally accessible.  Access to a large number of production modes and decay final states provides sensitivity to multiple Higgs couplings.  In the early stages of Higgs coupling measurements at the LHC, input from the Tevatron on $b \bar b$ final states at the low end of the allowed mass range may be useful, even if it provides only an upper bound.

Taking ratios of Higgs signal rates in different channels provides ratios of Higgs couplings-squared~\cite{Zeppenfeld:2000td}.  Adding a theoretical assumption, e.g., that the $HWW$ and $HZZ$ couplings are bounded from above by their SM value~\cite{Duhrssen:2004cv}, or that the Higgs does not have any significant non-SM decay modes~\cite{Zeppenfeld:2000td,Belyaev:2002ua,Lafaye:2009vr}, allows the individual Higgs couplings-squared to be extracted.  For a Higgs in the ``light'' allowed mass range of 114.4--145~GeV, ratios of the accessible couplings-squared are expected to be measurable at the 10--30\% level~\cite{Zeppenfeld:2000td}, and individual couplings-squared are expected to be measurable at the 20--70\% level~\cite{Duhrssen:2004cv}.  A careful treatment of theoretical and experimental uncertainties is very important; coupling extraction for $M_H = 120$~GeV has been studied with updated (2009) uncertainty estimates in Ref.~\cite{Lafaye:2009vr}.

A second key Higgs coupling measurement is the determination of the Lorentz structure of the Higgs coupling to $WW$ or $ZZ$ pairs.  The SM Higgs coupling has the special form $H V_{\mu} V^{\mu}$, arising specifically from the gauge-covariant derivative of the vev-carrying, weak-charged Higgs doublet.  In contrast, generic loop-induced couplings for a neutral scalar take the form $\phi V_{\mu\nu} V^{\mu\nu}$ for a CP-even scalar, or $\phi V_{\mu\nu} \widetilde V^{\mu\nu}$ for a CP-odd scalar, with $\widetilde V^{\mu\nu} = \epsilon^{\mu\nu\rho\sigma} V_{\rho\sigma}$.  These couplings can be distinguished experimentally using angular correlations of the forward tagging jets in weak boson fusion Higgs production or the four final-state fermions in $H \to VV$ decays.  For example, the azimuthal angle $\Delta \phi_{jj}$ of the forward tagging jets in weak boson fusion has a fairly flat distribution for the SM $H V_{\mu} V^{\mu}$ coupling, while for the CP-even (CP-odd) loop-induced vertex the distribution peaks at $\Delta \phi_{jj} \sim 0$, $\pi$ ($\pi/2$, $3\pi/2$)~\cite{Hankele:2006ma}.

\subsection{The Hierarchy Problem}

Despite the successes of the SM, we expect more than just the SM Higgs at LHC energies.  The argument for this new physics---based on the hierarchy problem associated with the SM Higgs---rests on the prejudice that extreme fine-tuning is implausible in nature.  This argument is thus weaker than the requirement of longitudinal vector boson scattering unitarity that guarantees a Higgs (or something to play its role) at LHC energies.

The Higgs mass-squared parameter $\mu^2$ in Eq.~(\ref{eq:V}) receives quantum corrections that depend quadratically on the high-scale cutoff of the theory.  For example, consider the radiative corrections to $\mu^2$ from a top quark loop: writing the renormalized parameter as $\mu^2 = \mu^2_0 + \Delta \mu^2$, where $\mu^2_0$ is the bare (high-scale) value, the one-loop diagram is given by
\begin{eqnarray}
	\Delta \mu^2 &\sim& - \int \frac{d^4 p}{(2\pi)^4} N_c {\rm Tr} \left[i \lambda_t \frac{i}{p_{\mu}\gamma^{\mu}} i \lambda_t \frac{i}{p_{\nu} \gamma^{\nu}} \right] \nonumber \\
	&=& - \frac{4 N_c \lambda_t^2}{(2 \pi)^4} \int \frac{d^4 p}{p^2},
\end{eqnarray}
where $\lambda_t$ is the $H t \bar t$ coupling and $p^{\mu}$ is the loop momentum, here taken large compared to $M_H$ and $m_t$.  Imposing a momentum cutoff $\Lambda$, the momentum integral diverges like $\Lambda^2$.  A full one-loop calculation of the top quark contribution yields
\begin{equation}
	\Delta \mu^2 = \frac{N_c \lambda_t^2}{16 \pi^2} \left[ -2 \Lambda^2 + 6 m_t^2 \ln(\Lambda/m_t) + \cdots \right].
\end{equation}
We measure $\mu^2 \sim - \mathcal{O}(M_{\rm EW}^2) \sim - (100~{\rm GeV})^2 = - 10^4~{\rm GeV}^2$.  Nature sets the bare parameter $\mu_0^2$ at the cutoff scale $\Lambda$.  If $\Lambda$ is the Planck scale $M_{\rm Pl} = 1/\sqrt{8 \pi G_N} \sim 10^{18}$~GeV, then $\Delta \mu^2 \sim -10^{35}$~GeV.  This huge correction must cancel against $\mu_0^2$ to yield the observed weak scale.

This is not an inconsistency in the theory---the SM is renormalizable, and the divergence can be absorbed into the bare parameter $\mu_0^2$ in the usual way.  The hierarchy problem comprises the implausibly huge top-down coincidence that $\mu_0^2$ and $\Delta \mu^2$ should cancel to 31 decimal places.  This coincidence is made even more implausible by the fact that loop contributions to $\Delta \mu^2$ arise from all the SM particles, introducing significant dependence of $\Delta \mu^2$ on several independent SM coupling parameters, and higher loop orders are also numerically significant, making the coupling parameter dependence highly nontrivial in form.\footnote{The most plausible proposed mechanism for such a fine-tuned cancellation is anthropic vacuum selection from an exponentially large multiverse.  However, a ``weakless'' universe, in which $\mu^2$ is pulled up to the Planck scale, could plausibly support intelligent observers~\cite{Harnik:2006vj}, thereby weakening the anthropic argument.}

To avoid the hierarchy problem, we need $|\Delta \mu^2| \sim (100~{\rm GeV})^2$.  This is achieved in the SM if $\Lambda \sim 1$~TeV; i.e., if new physics that eliminates the quadratically-divergent contributions to the Higgs mass-squared parameter appears at the TeV scale.

Before considering solutions to the hierarchy problem, let's ask the question whether the SM can be valid all the way to the Planck scale.  This is nontrivial due to radiatively-induced instabilities in the Higgs self-coupling, $\lambda$ in Eq.~(\ref{eq:V}), at high energy scales for too-large or too-small weak-scale values of $\lambda$.  Radiative corrections from top quark loops make $\lambda$ run smaller with increasing energy scale, while loops of the Higgs itself cause $\lambda$ to run larger with increasing energy scale.  If $\lambda$ is too large at the weak scale, the Higgs loop dominates and $\lambda$ blows up at some intermediate scale (this is called a Landau pole).  If $\lambda$ is too small at the weak scale, the top loop dominates and $\lambda$ runs negative at some intermediate scale, destabilizing the vacuum.  Either of these imply new physics at or below the intermediate scale.  Because $\lambda$ controls the physical Higgs mass [see Eq.~(\ref{eq:MH})], this allows conclusions about high-scale possibilities to be drawn based on the physical Higgs mass.

The range of Higgs masses allowed if the SM is to be valid up to various intermediate scales was studied in Ref.~\cite{Hambye:1997ax} (for a review see also Ref.~\cite{Quigg:2009vq}).  Using $m_t = 175$~GeV and $\alpha_s(M_Z) = 0.118$, Ref.~\cite{Hambye:1997ax} found a window in which the SM is perturbative and stable (but terribly fine-tuned) up to the Planck scale for $M_H \simeq 134$--180~GeV.  (The lower limit decreases if a smaller top quark mass is used; I will revisit this in the context of high-scale SUSY in the next section.)  Part of this window, $M_H > 145$~GeV, is already excluded by the LHC~\cite{LP11SMHiggs}.  Thus the LHC measurement of the Higgs mass itself has the potential to refute or preserve this distasteful possibility.

There are two main classes of solutions to the hierarchy problem, which I discuss in the next sections.  The first is supersymmetry (SUSY).  SUSY relates $\mu^2$ to a fermion mass, which receives radiative corrections that depend only logarithmically on the high-scale cutoff $\Lambda$.  This relation guarantees a cancellation of the quadratically-divergent part of the Higgs mass radiative corrections between loops of SM particles and loops of their SUSY partners.  The second is compositeness.\footnote{I include extra-dimensional models like Randall-Sundrum in the compositeness class via the AdS/CFT duality.}  In this class of models the Higgs is a bound state of new fundamental fermions, held together by a new force that becomes confining at the TeV scale.  Above the TeV scale there are no fundamental scalars, and thus no hierarchy problem.

%%%%%%%%%%%%%%%%%%%%%%%%%%%%%%%%%%%%%%%%%%%%
\section{FRAMEWORKS FOR PHYSICS BEYOND THE STANDARD MODEL I: SUPERSYMMETRY}
\label{sec:SUSY}

\subsection{The MSSM}

The Minimal Supersymmetric Standard Model (MSSM) scarcely needs an introduction; for a pedagogical review and references, see, e.g., Ref.~\cite{Martin:1997ns}.  MSSM signals depend heavily on the mass spectrum of SUSY particles, which determines the kinematics and identities of the particles produced in cascade decay chains.  The expected signatures are generically:
\begin{itemize}
\item Jets plus missing transverse energy (MET): squarks and/or gluinos are produced with large cross sections through QCD interactions and decay via a cascade to an invisible lightest SUSY particle (LSP);
\item Leptons plus MET: sleptons and weak gauginos are produced via electroweak interactions; the production couplings are smaller, but these colourless particles are typically lighter than squarks and gluinos due to renormalization group running (see below);
\item Third-generation fermions plus MET: renormalization group running tends to drive the top and bottom squarks and tau slepton lighter than the others due to the effect of their relatively large Yukawa couplings, leading to their predominance in cascade decays;
\item Photons plus MET: if SUSY is broken at an intermediate scale, the gravitino can be the LSP, leading to cascade decays to the next-to-lightest SUSY particle (NLSP; typically a bino) followed by bino decays to photon plus gravitino.
\end{itemize}

The mass spectrum of SUSY particles is determined by the parameters at the high-energy SUSY-breaking scale together with the renormalization group running down to the weak scale.  There are several generic features that appear in many specific models:
\begin{itemize}
\item Squarks (and separately sleptons) typically have a common mass at the high scale in order to avoid flavour problems.  Their masses are therefore degenerate except for splittings caused by renormalization group effects;
\item Renormalization group running typically makes the coloured particles significantly heavier at the weak scale than the colourless particles;
\item Contributions to renormalization group running from Yukawa couplings decrease the sfermion mass at the weak scale, leading to lighter third-generation sfermions;
\item  Left-right sfermion mixing is large when the Yukawa coupling is large, leading to a mass splitting that further decreases the mass of the lighter third-generation sfermions;
\item Heavy top squarks pull up the mass of the lightest MSSM Higgs boson $h^0$, but too-heavy top squarks reintroduce fine tuning.  Top squarks at 1~TeV are probably all right; at 2~TeV the fine-tuning is starting to become a bit uncomfortable;
\item The gluino mass is an independent parameter from the squark masses;
\item The gluino:wino:bino mass ratios are fixed by the SUSY-breaking mechanism.  In Minimal Supergravity (mSUGRA) and gauge-mediated SUSY breaking their weak-scale ratios are 7:2:1, while in anomaly-mediated SUSY breaking they are 8.3:1:2.8, leading to a wino LSP with a nearly degenerate chargino NLSP and difficult-to-detect signatures with little missing energy.
\end{itemize}

Supersymmetric models definitely predict a Higgs boson.  The lightest Higgs $h^0$ of the MSSM tends to have couplings very similar to those of the SM Higgs.  The MSSM with top squarks below 1~TeV \emph{requires} $M_{h^0} \lesssim 135$~GeV~\cite{Carena:2002es}.  The extra Higgs states of the MSSM, $H^0$, $A^0$, and $H^{\pm}$, can be heavy and tend to be nearly degenerate; their production cross sections and decays depend strongly on the MSSM Higgs sector parameter $\tan\beta$.

SM Higgs searches are thus very relevant in the MSSM.  The light SM-like Higgs $h^0$ is predicted to lie in the mass range where $gg \to H \to \gamma\gamma$ is the most sensitive channel for the SM Higgs and where the LHC does not yet have exclusion sensitivity.  Furthermore, $h^0$ can ``hide'' when $A^0$ is relatively light (100--200~GeV) and $\tan\beta \gtrsim 10$, mainly due to mixing between the two CP-even Higgs mass eigenstates that can suppress the $gg \to h^0$ cross section while simultaneously suppressing the $h^0$ branching ratio to $\gamma\gamma$ by enhancing its partial width to the dominant $b \bar b$ decay mode~\cite{Carena:2011fc}.  Fortunately, this difficult region of the $M_A$--$\tan\beta$ plane has subsequently been mostly excluded by direct searches for $A^0$, $H^0 \to \tau\tau$~\cite{tautau:LP11}.  This illustrates the importance of the interplay among different search channels in specific models, which will become increasingly significant as LHC data accumulates.

\subsection{A Heavier Higgs in SUSY Models}

Supersymmetry is compatible with a SM-like Higgs above 135~GeV, but making the Higgs heavier requires some modifications of the simplest MSSM.

One avenue to raise the SM-like Higgs mass is to add a Higgs coupling to a new singlet scalar supermultiplet $S$ of the form $\lambda S H_u H_d$, where $H_u$ and $H_d$ are the two Higgs doublets of the MSSM (this extension is called the next-to-minimal supersymmetric SM, or NMSSM).  To significantly raise the SM-like Higgs mass, a large value of $\lambda$ is needed; radiative corrections to $\lambda$ then cause it to blow up at an intermediate scale (a Landau pole).  This need not be a problem if the Higgs fields are composites of new strong dynamics that appears above the Landau pole.  Such a model, called the Supersymmetric Fat Higgs model~\cite{Harnik:2003rs} (``fat'' referring to the non-pointlike nature of the Higgs), can yield a SM-like Higgs with mass of order 200--450~GeV.  While most of this mass range is now excluded for the SM Higgs, the Fat Higgs could still be viable through suppressed production due to mixing among the mass eigenstates and/or suppressed branching ratios to $WW$ and $ZZ$ channels due to new decays to SUSY particles.  A detailed study of the LHC constraints on the parameter space has not yet been done.

A second avenue to raise the SM-like Higgs mass within the MSSM is to push up the masses $M_{\tilde t}$ of the top squarks.  This increases the loop-induced contribution to the $h^0$ mass, which grows with $\ln(M_{\tilde t}/m_t)$; however, it simultaneously increases the fine-tuning, which grows with $M_{\tilde t}^2$.  The most extreme possibility is to push all the SUSY particles, and the extra Higgs bosons of the MSSM, up to the GUT or Planck scale.  This ``Supersplit Supersymmetry''~\cite{Fox:2005yp}\footnote{Or more respectably, ``High-Scale SUSY''; the original Supersplit Supersymmetry paper is believed to have been a brilliant April Fool's joke.} makes a sharp prediction for the Higgs mass based on the high-scale supersymmetric boundary conditions that relate the Higgs quartic couplings to the SM gauge couplings.  The resulting Higgs mass is $141 \pm 2$~GeV for $\tan\beta \gtrsim 5$, falling to $128 \pm 2$~GeV for $\tan\beta \simeq 1$~\cite{Hall:2009nd}, using $m_t = 173.1$~GeV and putting the rest of the SUSY particles at $10^{14}$~GeV.  This is the same phenomenon as the perturbative and stable window for the fine-tuned SM, but with the Higgs quartic coupling fixed at the high scale by SUSY.  Note the reduced lower limit of this window compared to the $\sim 134$~GeV quoted in Ref.~\cite{Hambye:1997ax} due to the lower top quark mass value used.  A recent careful reanalysis~\cite{Giudice:2011cg} in Supersplit Supersymmetry using full two-loop renormalization group equations predicts a Higgs mass in the range (130--141)$\pm 2$~GeV, with the uncertainty mostly due to the current 0.9~GeV uncertainty in the top quark mass.

A less extreme (but still very fine-tuned) possibility is ``Split Supersymmetry''~\cite{ArkaniHamed:2004fb,Giudice:2004tc}, in which all scalar SUSY partners (and the extra MSSM Higgs states) are pushed up to an intermediate scale but the Higgsinos and gauginos are kept at the TeV scale in order to retain gauge coupling unification and a viable dark matter candidate.  The predicted light Higgs masses are several GeV heavier than in the Supersplit case due to radiative corrections from the Higgsinos and gauginos~\cite{Giudice:2004tc,Giudice:2011cg}.\footnote{Keeping only the gauginos light and requiring the right amount of wino dark matter yields $M_{h^0} \simeq 142 \pm 2$~GeV~\cite{Unwin:2011ag}, more similar to the prediction in Supersplit Supersymmetry.}  In Split SUSY, the gluino decays only via three-body processes involving a very heavy virtual squark.  This leads to displaced gluino decay vertices or metastable gluinos which hadronize into charged massive particles; these exotic signatures could be key to constraining the scenario~\cite{Alves:2011ug}.

%%%%%%%%%%%%%%%%%%%%%%%%%%%%%%%%%%%%%%%%%%%%%%
\section{FRAMEWORKS FOR PHYSICS BEYOND THE STANDARD MODEL II: COMPOSITE MODELS}
\label{sec:composite}

Models of compositeness trace electroweak symmetry breaking to new strong dynamics at or above the TeV scale.  These models fall into three broad classes: technicolor (or Higgsless models), composite-Higgs models, and little Higgs models.

\subsection{Technicolor}

Technicolor~\cite{Technicolor} is a class of strongly-coupled theories that become confining at the TeV scale.  These models contain no Higgs boson per se.  The Goldstone bosons ``eaten'' to form the longitudinal polarization states of the $W$ and $Z$ are bound states of the strongly-coupled dynamics, referred to as techni-pions in analogy to QCD.  

Strongly-coupled theories are hard to calculate in reliably; for this reason early technicolor analyses used the analogy with QCD by simply scaling up QCD measurements to the TeV scale.  QCD-like technicolor was thoroughly excluded by LEP-I precision electroweak data, and as a result the class of theories was largely ignored until the recent development of new techniques for computing in strongly-coupled theories.  These new techniques include calculations directly in the strongly-coupled gauge theory, lattice calculations, and calculations in five-dimensional gravitational theories believed to yield equivalent dynamics to four-dimensional conformal field theories via the AdS/CFT correspondence~\cite{Maldacena:1997re}.  

The AdS/CFT approach can be used to construct phenomenologically-useful four-dimensional ``Higgsless'' effective theories via deconstruction~\cite{Hill:2000mu}.  These effective theories typically contain spin-one techni-rho resonances ($W^{\prime}$ and $Z^{\prime}$ states) below the TeV scale; exchange of these resonances partially unitarizes longitudinal $WW$ scattering without a physical Higgs and pushes up the scale at which the full strong dynamics must appear.  These resonances typically have suppressed couplings to SM fermions and instead decay predominantly into $WW$ or $WZ$.  Depending on the global symmetries in the strongly coupled theory, there can also be physical (uneaten) techni-pions in the low-energy theory.  These techni-pions can be searched for in the $gg \to \pi \to \gamma\gamma$, $\tau\tau$ channels at the LHC if the theory includes coloured techni-fermions, and the SM Higgs search limits are already putting strong constraints on some models~\cite{Chivukula:2011ue}.

It is generically difficult to generate a large enough top quark mass in technicolor theories.  This has led to extensions such as topcolor-assisted technicolor~\cite{TC2}.  This scenario introduces a second new strong interaction felt only by the third-generation quarks which leads to a bound state of $\bar t_R Q_L$ with the quantum numbers of the Higgs doublet and a nonzero (but relatively small) vacuum expectation value $f = v \sin\omega$, where $\sin\omega$ is expected to be $\sim$0.2--0.5.  This ``top-Higgs'' doublet couples strongly to the top quark and is responsible for most of its mass.  The physical top-Higgs particle $H_t$ couples only to $t \bar t$, $WW$, and $ZZ$ at tree level; its couplings to $WW$ and $ZZ$ are suppressed relative to the SM Higgs by the $\sin\omega$ factor defined above, while its $t \bar t$ coupling is enhanced by $1/\sin\omega$, leading to a significantly enhanced gluon fusion production cross section.  The typical top-Higgs mass is $M_{H_t} \lesssim 2 m_t$ for dynamical top mass generation in topcolor-assisted technicolor.  
The spectrum also contains an isospin triplet of top-pions $\Pi_t$ (actually mixtures of the Goldstone bosons of the top-Higgs doublet and those of the technicolor electroweak symmetry breaking sector); their masses are constrained mainly by Tevatron limits on the exotic top quark decay $t \to \Pi_t^+ b$. 

Applying the LHC Higgs search to $gg \to H_t \to WW$, $ZZ$ excludes most of the topcolor-assisted technicolor parameter space~\cite{Chivukula:2011dg}.  Only a sliver of parameter space with $H_t$ between 300 and 350~GeV is still allowed, and this only when the top-pions are lighter than $m_t$ so that $H_t \to \Pi_t\Pi_t$ decays can suppress the top-Higgs branching ratios to $WW$ and $ZZ$.  Direct searches for $\Pi_t^0 \to \gamma\gamma$ may close this window, killing off a major attractive top quark mass generation mechanism in technicolor theories.  Still viable is the top seesaw~\cite{top-seesaw}, which leads to a significantly heavier top-Higgs that is more difficult to detect at the LHC.

\subsection{Composite Higgs Models}

Generating fermion masses and satisfying electroweak precision constraints in strongly-coupled models is generally much more straightforward when a physical Higgs boson is present in the spectrum.  The prototype for such composite Higgs theories is the Randall-Sundrum model~\cite{Randall:1999ee}.  This model contains a warped fifth dimension in which the SM fields propagate, bounded by a ``brane'' at each end.  The warping causes a Planck-scale cutoff on one brane (the Planck or UV brane) to be redshifted down to a much lower scale on the other brane (the TeV or IR brane).  The physical Higgs particle has a fifth-dimensional profile localized on or near the IR brane, keeping the cutoff for Higgs radiative corrections low and thereby solving the hierarchy problem.  The top quark must also be localized fairly close to the IR brane in order for it to have a large enough wavefunction overlap with the Higgs to generate the large top mass.  The Kaluza-Klein excitations of the SM fields in the five-dimensional theory are interpreted as bound states of the strongly-coupled gauge theory in the four-dimensional picture.  In this picture the Higgs is composite and the top quark is partly composite.

The physical signatures are the Kaluza-Klein excitations of the $Z$, $W$, and gluon; because of the localization of the top quark, these typically have enhanced couplings to the right-handed top and decay preferentially to $t \bar t$ pairs.  The Kaluza-Klein excitations of quarks can be pair produced via QCD interactions or singly produced via $qW$ or $qZ$ fusion, with decays back to $qW$ and $qZ$.

\subsection{Little Higgs Models}

Electroweak precision constraints remain a major concern for any model of strongly-coupled new physics at the TeV scale.  The new particles generically contribute to tree-level SM processes like $f \bar f \to f \bar f$, which have been measured with high precision.  New features in the electroweak and top-quark sectors are generically constrained by the oblique parameters $S$ and $T$ and by the $b \bar b$ fraction of hadronic $Z$ decays, $R_b$.  These constraints generically push the scale of the new physics---and the effective cut-off scale for the Higgs mass quadratic divergence---above the ``natural'' scale of $\lesssim 1$~TeV, requiring complicated model-building to keep the scale as low as even a few TeV.  This problem is known as the ``little hierarchy'' and is the target of ``little Higgs'' models~\cite{LH,ArkaniHamed:2002qy}.  

Little Higgs models make the Higgs a pseudo-Goldstone boson of a global symmetry that is broken only ``collectively'' by at least two different operators.  This allows quadratically-divergent contributions to the Higgs mass to be generated only at the two-loop level.  Naturalness at two loops requires a cutoff around 10~TeV, allowing the strong dynamics to be pushed up to this higher scale.  In practice, the cancellation of the one-loop Higgs mass radiative correction is accomplished by a minimal set of new particles at the TeV scale to cancel only the most important top-quark and SU(2) gauge boson contributions.  The ``Littlest Higgs'' model~\cite{ArkaniHamed:2002qy} is a prototypical example in which the phenomenology has been thoroughly studied; it contains TeV-scale partners for the top quark, the SU(2) gauge bosons, the Higgs itself, and (in some versions) the U(1) gauge boson.

The top-partner $T$ in the Littlest Higgs model is an SU(2)-singlet vectorlike quark, produced in pairs via QCD or singly via $bW \to T$ with decays to $bW$, $tZ$, and $tH$.  The neutral-current decays distinguish it from a sequential fourth-generation $t^{\prime}$. The gauge partners $W_H$ and $Z_H$ come from the breaking of SU(2)$\times$SU(2)$\to$SU(2)$_L$ at the TeV scale.  They couple to left-handed fermions via the usual SU(2) gauge generators $T^{\pm,3}$, but with coupling strength $g \cot\theta$, where $\theta$ is a free mixing parameter determined by the relative strengths of the gauge couplings of the two original SU(2) groups.  These gauge partners decay to SM fermions and also to $HV$ and $VV$ (where $V = W$ or $Z$; this latter decay makes their signatures similar to techni-rhos).  Tree-level exchange of these gauge partners in SM $f \bar f \to f \bar f$ processes puts rather tight constraints on their mass and coupling, tending to push them above 2~TeV~\cite{Csaki:2002qg}.  Phenomenology and LHC detection prospects for the $T$, $W_H$ and $Z_H$ have been studied in, e.g., Ref.~\cite{Azuelos:2004dm}.  Recent LHC searches for $W^{\prime} \to \ell \nu$~\cite{Aad:2011yg} are just beginning to probe the parameter space allowed by the electroweak precision constraints.

Electroweak precision constraints on little Higgs models tend to push the TeV-scale partners heavier than one would like to avoid fine tuning.  These constraints can be fully evaded by building the model with an extra parity, called ``$T$-parity'', under which the extra TeV-scale states are odd~\cite{Cheng:2003ju,Cheng:2004yc}.  This eliminates tree-level contributions to electroweak precision observables, loosening the constraints much like in SUSY.  It also leads to very different collider phenomenology: the $T$-odd particles are produced only in pairs, and decay to a lightest $T$-odd particle which can be a dark matter candidate.  Searches for the top-partner decaying to $t$ plus the invisible lightest $T$-odd particle are already being made at the LHC~\cite{toppart}.  

Finally, the Higgs search can have major implications for little Higgs models: the Higgs masses preferred by precision electroweak constraints together with minimization of fine-tuning can be quite high in some models---in the ``Littlest Higgs with $T$-parity''~\cite{Cheng:2004yc} the preferred Higgs masses are above 300~GeV~\cite{Hubisz:2005tx}.  The impact of the new LHC results on little Higgs models have yet to be systematically explored.

%%%%%%%%%%%%%%%%%%%%%%%%%%%%
\section{SUMMARY AND OUTLOOK}
\label{sec:outlook}

Early LHC data is already having impacts on a wide range of models.  The major task for phenomenologists at this moment is to incorporate the new LHC exclusions into the model ``landscape'' and to understand the resulting implications on model viability and fine-tuning.  In this talk I have focused mostly on the present and future impact of the LHC Higgs search because it has implications for a very wide range of models for new physics.

None of the major classes of new-physics models can be considered fully dead yet, but some are starting to be tightly constrained (e.g., topcolor-assisted technicolor).  The outcome of the LHC Higgs search with 5--10~fb$^{-1}$ will be extremely important.

\begin{acknowledgments}
I thank the organizers of Physics in Collision for the invitation to speak and for putting together an exciting symposium in a beautiful environment. This work was supported by the Natural Sciences and Engineering Research Council of Canada.
\end{acknowledgments}

%% Create the reference section using BibTeX:
%%\begin{thebibliography}{9}   % Use for  1-9  references

\end{document}